\documentclass[12pt]{article}
\usepackage{epsfig}
\usepackage{pst-plot,url,epsf}

\newcommand{\beq}{\begin{equation}}
\newcommand{\eeq}{\end{equation}}
\newcommand{\bea}{\begin{eqnarray}}
\newcommand{\eea}{\end{eqnarray}}
\newcommand{\beas}{\begin{eqnarray*}}
\newcommand{\eeas}{\end{eqnarray*}}
\newcommand{\epm}{e^+e^-}
\newcommand{\ra}{\rightarrow}

\newcommand{\eemmbb}{e^+ e^- \ra \mu^+ \mu^- \bar{b} b}

\newcommand{\eeudmn}{e^+ e^- \ra u \bar{d} \mu^- \bar{\nu}_{\mu}}

\newcommand{\nn}{\nonumber}

\begin{document}
\vspace*{1.0cm}
\begin{center}
{\LARGE\bf {FACTORIZABLE CORRECTIONS TO\\ 
                    $\eemmbb$ \\
                   AT A LINEAR COLLIDER\footnote{Presented by T. Westwa\'nski at the XXIX International Conference of 
        Theoretical Physics ``Matter to the Deepest: Recent Developments 
        in Physics of Fundamental Interactions'',
        Ustro\'n, Poland, September 9--14, 2005.
        Based on work in collaboration with Fred Jegerlehner.\\
        Work supported in part by the Polish State Committee for 
        Scientific Research in years 2005--2007 as a research grant.}}}

\vspace*{2cm}
Karol Ko{\l}odziej$^{\rm 2}$ and Tomasz Westwa{\'n}ski\footnote{E-mails: kolodzie@us.edu.pl, twest@server.phys.us.edu.pl}\\
\vspace*{0.3cm}
{\small\it
Institute of Physics, University of Silesia,\\
ul. Uniwersytecka 4, PL-40007 Katowice, Poland}\\
\vspace*{2.5cm}
{\bf Abstract}\\
\end{center}
We discuss the standard model factorizable radiative corrections to
$\eemmbb$, which is one of the best detection channels of the low mass Higgs 
boson produced through the Higgsstrahlung mechanism at a linear collider. The discussion
includes the leading virtual and real quantum electrodynamics corrections due to
the initial state radiation, the one-loop electroweak factorizable
corrections to the on-shell Higgs{\-}strahlung reaction and to subsequent decays
of the $Z$ and Higgs boson, and the quantum chromodynamics corrections to the Higgs 
boson decay width into a $b\bar b$-quark pair. 

\section{Introduction}
If the Higgs boson exists, it will be most probably discovered
at the Large Hadron Collider. However, the precise study of its production and decay 
properties, which will be necessary in order to establish the Higgs mechanism as 
the mechanism of the electroweak (EW) symmetry breaking of the standard model (SM),  
can be best performed
in a clean experimental environment of $\epm$ collisions at a future
International Linear Collider (ILC) \cite{ILC}.

One of the main production
mechanisms of the SM Higgs boson at the ILC is the Higgsstrahlung reaction 
\bea
\label{eeZH}
                      e^+e^- \rightarrow  Z H.
\eea
Reaction (\ref{eeZH}) dominates the Higgs boson production at low
energies, as its cross section scales like $1/s$, at same time when cross
sections of the Higgs boson production through the $WW$ and $ZZ$ boson fusion mechanisms
grow with the centre of mass (CM) energy as $\ln \; (s/m^{2}_{H})$.

If the Higgs boson is light, with mass between the present $\rm {95 \% \; CL}$
lower limit of $m_H=114.4$ GeV \cite{LEPdir} and an upper value of, say, $m_H=140$~GeV,
then it will decay dominantly into a pair of $b\bar b$ quarks.
As the Z boson of reaction (\ref{eeZH}) decays into a fermion-antifermion pair too, one
actually observes the Higgsstrahlung through reactions with four
fermions in the final state. Precise measurements of such reactions at the ILC
should be confronted with at least equally precise SM predictions, which obviously
must include radiative corrections. Because of six external particles and large
number of the contributing Feynman diagrams, 
calculation of the complete EW $\mathcal{O}(\alpha)$
corrections to reactions with four-fermion final states is very complicated.
Problems encountered in the first attempt
of such a complete calculation were described in \cite{Vicini}.
Substantial progress in a full one-loop calculation for
$\eeudmn$ was reported by the GRACE/1-LOOP team \cite{GRACE}
and quite recently such the calculation 
for the charged current reactions
$e^{+}e^{-} \ra \nu_{\tau} \tau^{+} \mu^{-} \bar \nu_{\mu}$ and
$e^{+}e^{-} \ra u \bar d s \bar c$, which are relevant for the $W$-pair production at
the ILC, has been accomplished in \cite{DDRW}.
However, there is no calculation of the complete EW $\mathcal{O}(\alpha)$ corrections
to the neutral-current $\epm \ra 4$~fermion processes available at the moment.

\vspace*{0.5cm}
\begin{figure}[!h]
\vspace{110pt}
\includegraphics{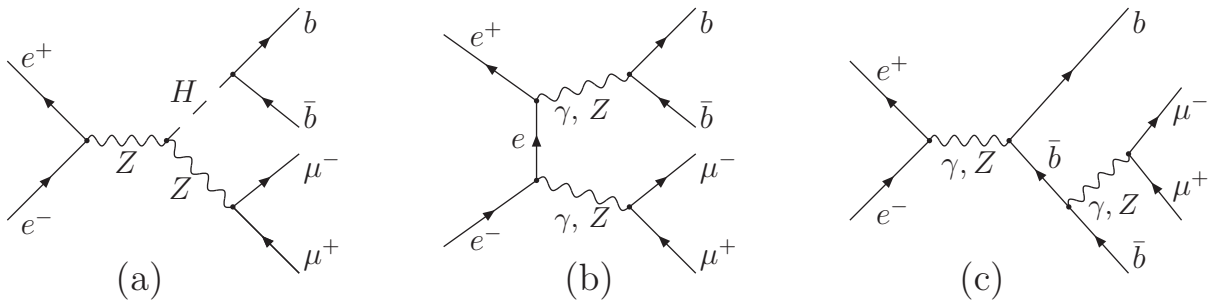}
\vspace{-0.2 cm}
\caption{Examples of Feynman diagrams of reaction (\ref{bmmb}):
(a) the double resonance `signal', (b) and (c) `background'
diagrams.}
\label{fig:diags}
\end{figure}
\vspace*{0.4cm}
Therefore, in the present lecture, we will discuss an alternative approach
to the problem of improving precision of the SM predictions for the Higgs boson
production and decay at the ILC through the Higgsstrahlung mechanism which
has been originally proposed in \cite{JKW} and recently accomplished in 
\cite{JKW2}. We will concentrate on the reaction
\bea
\label{bmmb}
\eemmbb,
\eea
which is one of the best detection channels of (\ref{eeZH}).
Typical examples of the Feynman
diagrams of reaction (\ref{bmmb}) are depicted in Fig.~\ref{fig:diags}. 
The Higgsstrahlung `signal' diagram is shown in Fig.~\ref{fig:diags}a,
while the diagrams in Figs.~\ref{fig:diags}b and \ref{fig:diags}c represent 
typical `background' diagrams. Altogether,
in the unitary gauge and with the neglect of the Higgs boson coupling to
electrons, there are 34 Feynman diagrams
which contribute to (\ref{bmmb}) in the lowest order of SM.
The lowest order SM cross section of reaction (\ref{bmmb}), as
well as of the corresponding bremsstrahlung reaction
\bea
\label{bmmbg}
\eemmbb\gamma,
\eea
which receives contribution from 236 Feynman diagrams, can be computed with 
a program {\tt ee4fg} \cite{ee4fg}.
On the basis of {\tt ee4fg}, we have written a dedicated program
{\tt eezh4f} that includes factorizable 
radiative corrections to (\ref{bmmb}) which will be discussed in Section 2.

\section{Factorizable radiative corrections}

We include the factorizable radiative corrections into the cross section of
(\ref{bmmb}) according to the following master formula \cite{JKW2}
\bea
\label{totalcs}
\int{\rm d}\sigma &=& \int{\rm d}\sigma_{\rm Born+r.m.} +
\int_{E_{\gamma}<E_{\rm cut}}
                 {\rm d}\sigma_{\rm virt + soft, univ.}^{\rm QED\;ISR}
		 \\ \nn
&+&\int_{E_{\gamma}>E_{\rm cut}}{\rm d}\sigma_{\rm hard}^{\rm QED\;ISR}
+\int{\rm d}\sigma_{\rm virt, finite}^{\rm EW\;DPA}.
\eea
The first integrand on the right hand side of (\ref{totalcs}) is the
effective Born cross section, calculated with the whole set of the Feynman
diagrams of (\ref{bmmb}),  in which we have included
the bulk of quantum chromodynamics (QCD) correction to the decay of Higgs boson 
into a $b\bar b$-quark pair that can be mapped into a running $b$-quark mass. 
This means in practise that 
the constituent $b$-quark mass that is used in the calculation 
is replaced in the Higgs--$b\bar b$ Yukawa coupling with the running mass, 
which is about a factor 3 smaller. 

The second integrand on the right hand side of (\ref{totalcs})
combines the universal infra-red (IR) singular part of the
$\mathcal{O}(\alpha)$ virtual QED correction to the on-shell $Z$--Higgs production
process (\ref{eeZH}) with the soft bremsstrahlung correction to
(\ref{bmmb}), integrated up to the soft photon energy cut $E_{\rm cut}$.
It can be written as 
\bea
\label{QEDvs}
{\rm d}\sigma_{\rm virt + soft, univ.}^{\rm QED\;ISR}=
{\rm d}\sigma_{\rm Born+r.m.}C_{\rm virt + soft, univ.}^{\rm QED\;ISR},
\eea
with the correction factor $C_{\rm virt + soft, univ.}^{\rm QED\;ISR}$ given by
\bea
C_{\rm virt + soft, univ.}^{\rm QED\;ISR}=
\frac{e^2}{2\pi^2} & &\hspace*{-0.5cm}\left[\left(\ln\frac{s}{m_e^2}-1\right)
          \ln\frac{2E_{\rm cut}}{\sqrt{s}}
         +\frac{3}{4}\ln\frac{s}{m_e^2} \right].
\label{CQEDvs}
\eea
In (\ref{CQEDvs}), $e$ is the electric charge that is given in terms
of the fine structure constant in the Thomson limit $\alpha_0$,
$e=\left(4\pi\alpha_0\right)^{1/2}$. Note that the running $b$-quark mass
correction is included in (\ref{QEDvs}). For the sake of consistency,
the same modification of the
$b$-quark mass in the Higgs--$b\bar b$ Yukawa coupling is done also in the third 
term on the right hand side of (\ref{totalcs})
that represents the initial state hard bremsstrahlung contribution,
{\em i.e.}
the cross section of reaction (\ref{bmmbg}) with the hard photon emitted
from the initial state electron or positron. The integral  over the hard 
photon phase space is performed over the full angular range of the photon momentum
and from the minimum value of
its energy $E_{\rm cut}$ to the very kinematical limit. 

Finally, the last integrand on the right hand side of Eq.~(\ref{totalcs}),
${\rm d}\sigma_{\rm virt, finite}^{\rm EW,\;DPA}$, is
the IR finite part of the virtual EW $\mathcal{O}(\alpha)$ correction
to reaction (\ref{bmmb}) in the double pole approximation (DPA). 
It can be written in the following way
\bea
\label{DPA}
{\rm d}\sigma_{\rm virt, finite}^{\rm EW,\;DPA}=
\frac{1}{2s}\left\{
\left|{M^{(0)}_{DPA}}\right|^2 {C_{\rm QED}^{\rm non-univ.}}
+ 2{\rm Re}\left({M^{(0)^*}_{DPA}\delta M_{DPA}}\right)\right\}
{\rm d}\Phi_{4f},
\eea
where $M^{(0)}_{DPA}$  and $\delta M_{DPA}$ are the lowest order matrix element 
and the one-loop correction, respectively, in the DPA. They
are calculated with  the projected four momenta $k_i$, $i=3,...,6$, of
the final state particles, except for denominators of the $Z$ and Higgs
boson propagators. The projected four momenta are obtained from
the four momenta $p_i$, $i=3,...,6$, of reaction (\ref{bmmb}) with, to some
extent arbitrary, a projection procedure which is described in \cite{JKW2};
${\rm d}\Phi_{4f}$ is the four-particle Lorentz invariant phase space element 
and $C_{\rm QED}^{\rm non-univ.}$ denotes the IR finite
non universal constant part of the $\mathcal{O}(\alpha)$ QED correction
that has not been taken into account in Eq.~(\ref{CQEDvs}). 
The interested reader is referred to \cite{JKW2} for detailed definitions
of $M^{(0)}_{DPA}$, $\delta M_{DPA}$ and $C_{\rm QED}^{\rm non-univ.}$,
as well as for the values of the physical input parameters used in the computation.

\begin{figure}[!h]
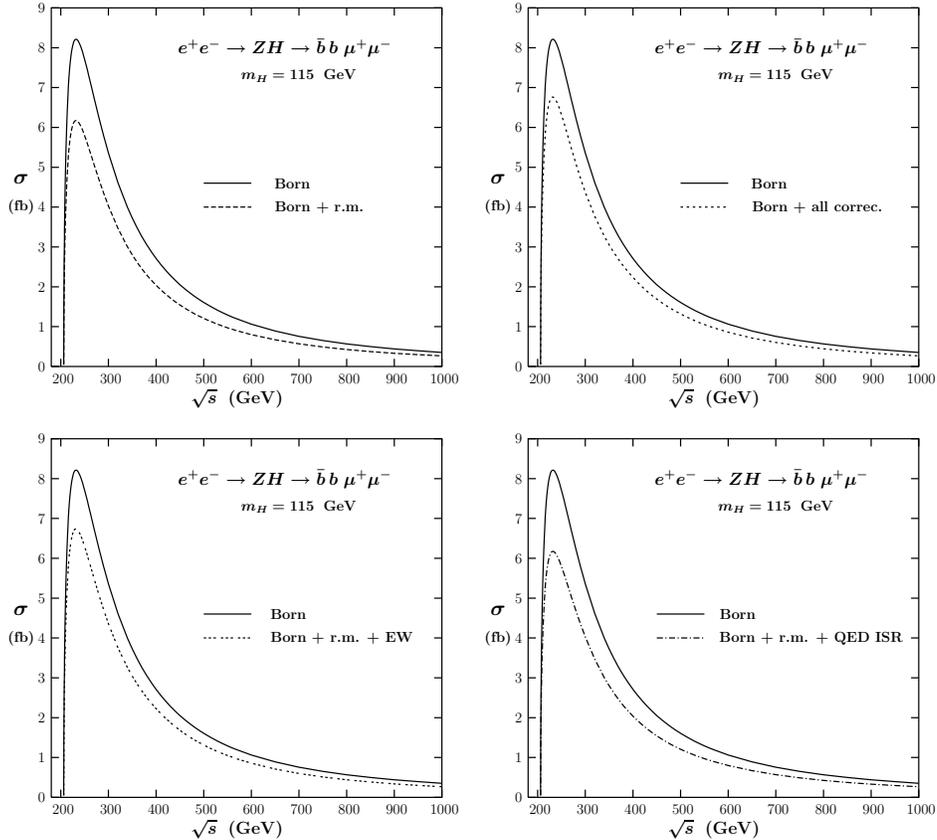

\begin{center}
\vspace*{0.5cm}
\setlength{\unitlength}{1mm}
\begin{picture}(35,35)(58,-50)
\rput(6.2,-4.3){\scalebox{0.46 0.46}{\epsfbox{fig2a.epsi}}}
\end{picture}
\begin{picture}(35,35)(9,-50)
\rput(4.,-4.3){\scalebox{0.46 0.46}{\epsfbox{fig2b.epsi}}}
\end{picture}
\vfill
\begin{picture}(35,35)(58,-50)
\rput(6.2,-6.5){\scalebox{0.46 0.46}{\epsfbox{fig2c.epsi}}}
\end{picture}
\begin{picture}(35,35)(9,-50)
\rput(4.,-6.5){\scalebox{0.46 0.46}{\epsfbox{fig2d.epsi}}}
\end{picture}
\end{center}
\vspace*{3.7cm}
\caption{The `signal' total cross section of reaction (\ref{bmmb})
         including different classes of corrections $\delta_{\rm cor.}$
         in the NWA as a function of the CMS energy for $m_H=115$~GeV.} 
\label{fig:sig}
\end{figure}

Numerical effects of the corrections described above are illustrated in
Fig.~\ref{fig:sig}, where we plot the total signal cross section of
(\ref{bmmb}) in the narrow width  approximation (NWA) for the $Z$ and Higgs boson.
How the running $b$-quark mass
correction reduces the Higgsstrahlung signal is illustrated 
in the upper left corner of Fig.~\ref{fig:sig}, where  we plot
the cross section of (\ref{bmmb}) in the NWA
to the lowest order (solid line)
\bea
\label{bornNWA}
\sigma_{\rm Born}^{\rm NWA}=\sigma_{e^+e^-\ra ZH}^{(0)}\;
\frac{\Gamma_{Z\ra \mu^+\mu^-}^{(0)}}{\Gamma_Z^{(0)}}\;
\frac{\Gamma_{H\ra b\bar b}^{(0)}}{\Gamma_H^{(0)}},
\eea
and including the running $b$-quark mass correction (dashed line)
\bea
\label{bornmb}
\sigma_{{\rm Born + running}\;m_b}^{\rm NWA}=\sigma_{e^+e^-\ra ZH}^{(0)}\;
\frac{\Gamma_{Z\ra \mu^+\mu^-}^{(0)}}{\Gamma_Z^{(0)}}\;
\frac{\Gamma_{H\ra b\bar b}^{(0+{\rm r.m.})}}{\Gamma_H^{(0+\rm{r.m.})}},
\eea
as functions of the centre of mass (CM) energy. In Eqs.~(\ref{bornNWA}) and
(\ref{bornmb}), $\sigma_{e^+e^-\ra ZH}^{(0)}$ is the lowest cross section
of the on-shell Higgsstrahlung reaction (\ref{eeZH}), 
$\Gamma_{Z\ra \mu^+\mu^-}^{(0)}$ and $\Gamma_Z^{(0)}$ 
$\left(\Gamma_{H\ra b\bar b}^{(0)}\right.$ and $\left.\Gamma_H^{(0)}\right)$ 
are the lowest order
partial and total $Z$ (Higgs) boson widhts, and
$\Gamma_{H\ra b\bar b}^{(0+{\rm r.m.})}$ and  $\Gamma_H^{(0+\rm{r.m.})}$
are the partial and total Higgs boson widhts including the running $b$ quark
mass correction. 

\begin{figure}[!h]
\begin{center}
\vspace*{-0.4 cm}
\setlength{\unitlength}{1mm}
\begin{picture}(35,35)(58,-50)
\rput(7.2,-6){\scalebox{0.65 0.65}{\epsfbox{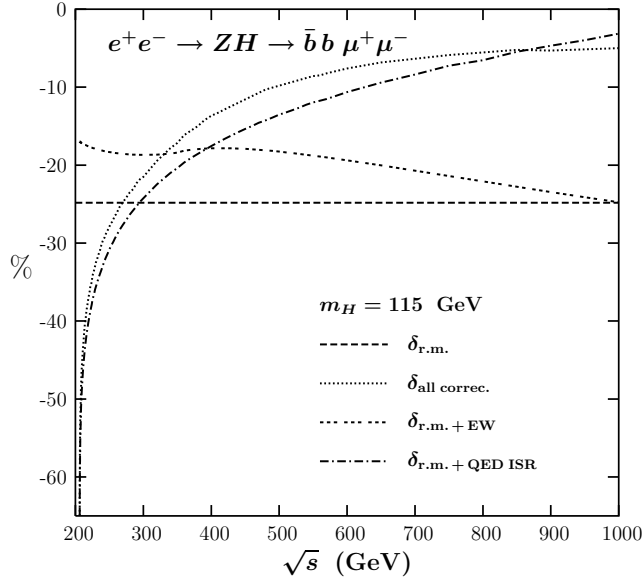}}}
\end{picture}
\end{center}
\vspace*{4.3 cm}
\caption{The relative corrections corresponding to those of 
         Fig.~{\ref{fig:sig}}.}
\label{fig:relcor}
\end{figure}

The numerical
effect of the QED initial state radiation (ISR) correction
is illustrated 
in the bottom right corner of Fig.~\ref{fig:sig}, where 
we plot the signal lowest order and
QED ISR corrected cross sections of (\ref{bmmb}) in the NWA. This correction is 
calculated in the structure function approach, as defined by Eq.~(42)
of \cite{JKW2}. Typical effects of the ISR, {\em i.e.} a shift in the position of
the maximum and a radiative tail, are obscured by the inclusion of
the running $b$ quark mass correction, which shifts the corrected cross section
downwards and makes its line shape narrower, due to the reduction of
the Higgs boson width.
The numerical effect of the EW and the running $b$ quark mass
corrections is shown in the bottom left 
corner of Fig.~\ref{fig:sig}. Finally, the combined effect of all the factorizable
corrections discussed is depicted in the upper right corner of Fig.~\ref{fig:sig}.
The corresponding relative corrections
\bea
\label{nwarc}
\delta_{{\rm cor.}}&=&\frac{\sigma_{{\rm
      Born+cor.}}-\sigma_{{\rm Born}}}{\sigma_{{\rm Born}}}
\eea
are plotted in Fig.~\ref{fig:relcor}.

\begin{table}
\label{tab1}
\caption{Branching ratios and the lowest order Higgs boson width.}
\begin{center}
\begin{tabular}{c r r r}
\hline
$m_H$  & BR$\left(H \ra b \bar b\right)$ & BR$\left(H \ra W^+ W^-\right)$ &
$\Gamma^{(0)}_{H}\hspace*{0.2cm} $ \\
  (GeV)  & (\%) \hspace*{0.7cm} & (\%) \hspace*{1.2cm} & (MeV) \\
\hline
115 & 81.29 \hspace*{0.5cm}  &  4.18  \hspace*{1.cm}  &   5.347  \\ 
130 & 67.53 \hspace*{0.5cm}  & 18.43  \hspace*{1.cm}  &   7.288 \\ 
150 & 29.59 \hspace*{0.5cm}  & 58.45  \hspace*{1.cm}  &  19.226  \\ 
160 &  5.28 \hspace*{0.5cm}  & 90.98  \hspace*{1.cm}  & 114.930   \\ 
\hline
\end{tabular}
\end{center}
\end{table}

\begin{figure}[!h]
\begin{center}
\vspace*{-0.4 cm}
\setlength{\unitlength}{1mm}
\begin{picture}(35,35)(58,-50)
\rput(7.2,-6){\scalebox{0.65 0.65}{\epsfbox{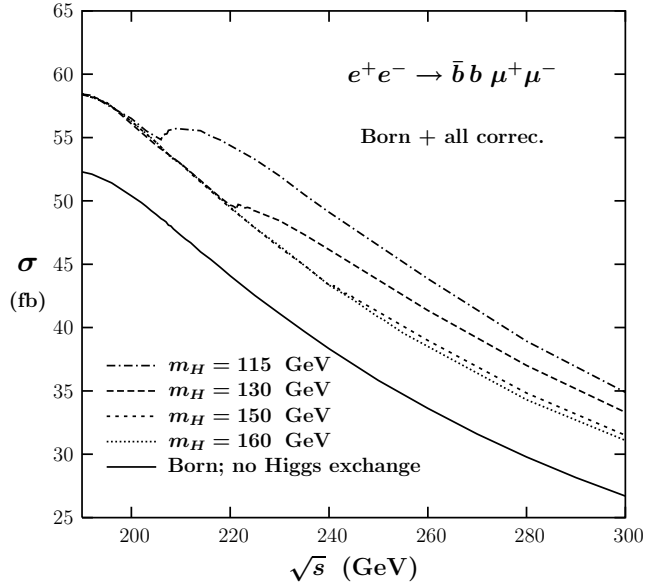}}}
\end{picture}
\end{center}
\vspace*{4.3 cm}
\caption{The total cross section of reaction (\ref{bmmb}) including
         all the corrections as a function of the CM energy for
         different values of $m_H$.}
\label{fig:corr}
\end{figure}

The total cross sections of reaction (\ref{bmmb}) including all the 
corrections as defined in Eq.~(\ref{totalcs}) are plotted in
Fig.~{\ref{fig:corr}} as functions of the CM energy for a few values of
the Higgs boson mass. The Higgs boson production
signal is clearly visible in the plots only for low values, $m_H=115$~GeV
and $m_H=130$~GeV, while it becomes hardly visible for
higher values of the Higgs boson mass, $m_H=150$~GeV and $m_H=160$~GeV. This effect 
is caused by a decrease in the
branching ratio ${\rm BR}(H \ra b \bar b)$ with the growing Higgs
boson mass. The values of the branching ratio ${\rm BR}(H \ra b \bar b)$,
together with ${\rm BR}(H \ra W^+ W^-)$ and the total lowest order SM Higss
boson width, the latter being used in the calculation, are collected in Table~1.
The Born cross section of reaction (\ref{bmmb}) without
the Higgs boson exchange (solid curve) is shown in Fig.~\ref{fig:corr}, too.
The plots of the corrected cross sections are shifted upwards with respect
to it, as they include
the initial hard bremsstrahlung contribution which is integrated over the
full photon phase space.

\section{Summary and outlook}
We have discussed  the  factorizable SM radiative corrections to reaction (\ref{bmmb}), 
which is one of the best detection channels of the light Higgs boson production
through the Higgsstrahlung mechanism at the
ILC. The leading virtual and real QED ISR corrections to
all the lowest order Feynman diagrams of reaction (\ref{bmmb}) have
been included. We have used the running $b$-quark
mass in the lowest order Higgs--$b\bar b$ Yukawa coupling
in order to include the bulk of QCD corrections to the Higgs boson decay
into the $b\bar b$-quark pair.
The complete electroweak $\mathcal{O} (\alpha)$ corrections to
the on--shell $Z$--Higgs production and to the $Z$ and Higgs decay widths have been
taken into account in the DPA. 
It has been illustrated how the corrections significantly reduce the Higgs
boson production signal cross section of (\ref{bmmb}) in the NWA.
Finally, we have shown how the Higgs boson production signal would be
visible in the CM energy dependence of the total cross section of (\ref{bmmb}).

\end{document}